\documentclass[11pt]{article}
\usepackage{graphicx}
\usepackage{amsmath}
\usepackage{amssymb}
\usepackage{amsxtra}
\def\Journal#1#2#3#4{#1 {\bf #2}, #3 (#4)}

\def\PLB{{ Phys. Lett.}  B}
\def\PRL{ Phys. Rev. Lett.}
\def\PRD{{ Phys. Rev.} D}

\def\JETPL{ JETP Lett.}

\def\CQG{ Class. Quantum Grav.}

\def\IJMPA{{ Int. J. Mod. Phys.}  A}

\def\BWP{ Bled Workshops in Physics}

\def\GeV{\,{\rm GeV}}
\def\TeV{\,{\rm TeV}}

\def\({\left(}
\def\){\right)}
\def\cm{{\,\rm cm}}

\def\beq{\begin{equation}}
\def\eeq{\end{equation}}
\def\bea{\begin{eqnarray}}
\def\eea{\end{eqnarray}}
\title{Beyond the Standard models of particle physics and cosmology}
\author{M.~Yu.~Khlopov\\
Institute of Physics,
Southern Federal University\\
Stachki 194,
Rostov on Don 344090, Russia\\ khlopov@apc.in2p3.fr}

\begin{document}
\maketitle

\begin{abstract}
The modern Standard cosmological model of inflationary Unvierse and baryosynthesis deeply involves particle theory beyond the Standard model (BSM). Inevitably, models of BSM physics lead to cosmological scenarios beyond the Standard cosmological paradigm. Scenarios of dark atom cosmology in the context of puzzles of direct and indirect dark matter searches, of clusters of massive primordial black holes as the source of gravitational wave signals and of antimatter globular cluster as the source of cosmic antihelium are discussed. 
\end{abstract}



\section{Introduction}\label{intro}
The basis of the modern Standard cosmological paradigm, involving inflation, baryosynthesis and dark matter as its neccessary basic elements, is related to new physics predicted in theory beyond the Standard model (BSM) of elementary particles (see e.g. Ref. \cite{DMRev} for review and reference). However, BSM models, reproducing the necessary basic elements of the modern cosmology, inevitably contain additional model dependent consequences that lead beyond the Standard cosmological scenario \cite{bled16}.  

Methods of cosmoparticle physics, studying fundamental relationship of cosmology and particle physics in the combination of its physical, astrophysical and cosmological signatures, involve such model dependent cosmological predictions to probe models of BSM physics and cosmological scenarios, based on them. \cite{book,newBook,bled}. 

 Here we show that BSM physics leads to cosmological scenarios accomplished by nontrivial deviations from the Standard cosmological model that deserve special interest in the context of the recent experimental progress.

We address a possibility of existence of stable double charged particles $O^{--}$ bound with primordial helium in neutral nuclear interacting O-helium dark atoms (Section \ref{darkatoms}) and consider advantages of this scenario to resolve puzzles of direct and indirect dark matter searches, as well as the open problems of OHe interaction with matter. We show that BSM physics of inflationary models that naturally leads to strong primordial inhomogeneities and to clusters of massive primordial black holes, in particular, is possibly reflected in the gravitational wave signal from massive black hole coalescence (Section \ref{pbhs}). We discuss in Section \ref{antiMatter} existence of antimatter stars in our Galaxy, originated from nonhomogeneous baryosynthesis in baryon asymmetrical Universe and reflected in cosmic antihelium fluxes, possibly detected by AMS02 \cite{amsAHe3,amsAHe}.
 
\section{Dark atom physics and cosmology}\label{darkatoms}
In the simplest case physics of dark matter is reduced to prediction by BSM model of new neutral elementary weakly interacting massive particle (WIMP). This type of prediction is beyond the standard model of elementary particles, but fits perfectly well the standard cosmological LambdaCDM paradigm. Supersymmetric (SUSY) models, predicting WIMP candidates, seemed to support this simple approach to dark matter physics. However negative results of experimental underground WIMP searches, as well as of collider searches for SUSY particles appeal to other possible BSM solutions for the dark matter problem. Possibly, SUSY physics and cosmology corresponds to superhigh energy scales as discussed in this Volume in \cite{ketov}.

In fact, the necessary conditions for dark matter candidates to be stable, satisfy the measured dark matter density and be decoupled from plasma and radiation at least before the beginning of matter dominated stage in no case demand these particle candidates to be weakly or superweakly interacting. Even nuclear interacting particles can play the same role due to decoupling of the gas of such particles from plasma and radiation before the end of radiation dominated stage. It gives rise to models of dark matter in the form of Strongly Interacting Massive Particles (SIMPs) \cite{Starkman,Wolfram,Starkman2,Javorsek,Mitra,Mack}. 

By definition dark matter should be 'dark', nonluminous, what seem to favor neutral elementary particles. However ordinary atomic matter is neutral but it is composite and consists of  electriclly charged particles (nuclei and electrons). In the same way O-helium dark atoms represent a specific example of composite SIMPs, in which hypothetical double charged $O^{--}$ particles are bound with primordial helium nuclei by ordinary Coulomb force \cite{I,FKS,KK,Norma,mpla,DDMRev}.
\subsection{OHe and $O$-nuclearites}
The main problem for hypothetical stable charged particles is their absence in the matter. If they do exist, they should be bound with ordinary matter and form anomalous isotopes. Severe experimental constraints on such isotopes, on anomalous hydrogen especially, seem to exclude a possibility for stable charged particles. However, if there exist stable particles with charge -2 in excess over corresponding particles with charge +2, such negatively charged particles are captured by primordial helium and form neutral OHe dark atom. There are various models, in which such stable -2 charged particles $O^{--}$ are predicted \cite{I,FKS,KK,Norma,mpla,DDMRev}. Moreover, if these particles possess electroweak SU(2) gauge charges, their excess can be equilibrated by electroweak sphaleron transitions with baryon excess, as it is the case in Walking Technicolor models \cite{KK}.

The general analysis of the bound states of single $O^{--}$ with nuclei was developed in a simple model \cite{Cahn,Pospelov,Kohri}.
For small nuclei the Coulomb binding energy is like in hydrogen atom
and is given by
\begin{equation}
    E_b=\frac{1}{2} Z^2 Z_O^2 \alpha^2 A m_p.
\end{equation}

For large nuclei $O^{--}$ is inside nuclear radius and the harmonic
oscillator approximation is valid for the estimation of the binding
energy
\begin{equation}
    E_b=\frac{3}{2}(\frac{Z Z_O \alpha}{R}-\frac{1}{R}(\frac{Z Z_O \alpha}{A m_p R})^{1/2}).
\label{potosc}
\end{equation}

Here $Z$ is the charge of nucleus, $A$ is its atomic number, $R$ is radius of nucleus, $Z_O=2$ is the charge of $O^{--}$, $m_p$ is the proton mass and $\alpha=1/137$ is the fine structure constant.
In the case of OHe $Z Z_O \alpha A m_p R \le 1$, what proves its
Bohr-atom-like structure (see \cite{mpla,DDMRev} for review and references).
However, the radius of Bohr orbit in these ``atoms"
\cite{I,KK} $r_{o} \sim 1/(Z_{O} Z_{He}\alpha m_{He}) \approx 2
\cdot 10^{-13} \cm $ is of
the order the size of
He nucleus. Therefore the corresponding correction to the binding energy due to
non-point-like charge distribution in He nucleus is significant.

$O^{--}$ particles are either elementary lepton-like states, or clusters of heavy $\Bar{U}$ quarks with charge -2/3 $\Bar{U}\Bar{U}\Bar{U}$, which have strongly suppressed QCD interaction. In the contrary to ordinary atoms OHe has heavy lepton-like core and nuclear interacting shell.

If multiple $O^{--}$ are captured by a heavy nucleus, the corresponding neutral bound system can acquire the form of $O$-nuclearites, in which negative charge of $O^{--}$ is compensated by posistive charge of protons in the nucleus \cite{GKV}. 
The energy of such a $O$-nuclearite is given by \cite{GKV}
\begin{equation}\label{E}
{\cal{E}}=-16 {\rm{
 MeV}} \cdot A  - \int d^3 r (n_\mathrm{p} -2n_O)V -\int d^3 r\frac{(\nabla{V})^2}{8\pi e^2} + {\cal{E}}_{\rm kin}^{O}\,.
\end{equation}
Here the first term is the volume energy of the atomic nucleus with atomic number A, next two terms describe the electromagnetic energy, and
\begin{equation}\label{E_kin}
{\cal{E}}_{\rm kin}^O = \displaystyle\int d^3 r \int\limits_0^{p_{{\rm F},O}^{}} \frac{p^2 dp}{\pi^2}\frac{p^2}{2m_O^{}}
\end{equation}
is the kinetic energy of the $O$-fermions of the mass $m_O$; $V = -e\phi$ is the potential well for the electron in the field of the positive charge ($e>0$, $\phi>0$) and on the other hand it is the potential well also for the protons in the field of the negative charge of $O$-particles. 

The most energetically favorable $O$-particle distribution inside the nucleus is that follows the proton one, fully compensating the Coulomb field. Thereby $O$-particles, if their number were $N_O\geq A/4$, would be re-distributed to minimize the energy, and finally the density of $O$ inside the atomic nucleus becomes $n_O = n_\mathrm{p}/2 = (n_\mathrm{p}^0/2)\:\theta(r-R)$ for $O$-nuclearite, that corresponds to $V = const$ for $r<R$. Excessive $O$-particles are pushed out.
\subsection{Cosmoparticle physics of OHe model}
 After the Standard Big Bang Nucleosynthesis (SBBN) $O^{--}$ charged particles capture $^4He$ nuclei in neutral 
OHe ``atoms'' \cite{I}. For the mass of $O^{--}$ $m_O \sim 1 \TeV$, $O^{--}$ abundance is much smaller than helium abundamce, so that He is in excess in such capture, making the abundance of frozen out free $O^{--}$ exponentially small.

The cosmological scenario of OHe Universe involves only one parameter
of new physics $-$ the mass of $O^{--}$. Such a scenario is insensitive to the properties of $O^{--}$ (except for its mass), since the main features of the OHe dark atoms are determined by their nuclear interacting helium shell. 

Before the end of radiation domination stage the rate of expansion exceeds the rate of energy and momentum transfer from plasma to OHe gas and the latter decouples from plasma and radiation. Then OHe starts to dominate at the Matter Dominated stage, playing the role of Warmer than Cold Dark Matter in the process of Large Scale Structure formation\cite{I,mpla}. This feature is due to conversion of small scale fluctuations in acoustic waves before OHe decoupling and to their corresponding suppression. However, the suppression of such fluctuations is not as strong as the free streaming suppression for few keV dark matter particles in Warm Dark matter models. 

In terrestrial matter OHe dark atoms are slowed down and cannot cause significant nuclear recoil in the underground detectors, making them elusive for detection based on nuclear recoil. The positive results of DAMA experiments (see \cite{DAMAtalk} for review and references) can find in this scenario a nontrivial explanation due to a low energy radiative capture of $OHe$ by intermediate mass nuclei~\cite{mpla,DMRev,DDMRev}. This explains the negative results of the XENON100 and LUX experiments. The rate of this capture is
proportional to the temperature: this leads to a suppression of this effect in cryogenic detectors, such as CDMS. 

OHe collisions in the central part of the Galaxy lead to OHe
excitations, and de-excitations with pair production in E0 transitions can explain the
excess of the positron-annihilation line, observed by INTEGRAL in the galactic bulge \cite{DMRev,DDMRev,KK2,CKWahe}. Due to the large uncertainty of DM distribution in the galactic bulge this interpretation of the INTEGRAL data is possible in a wide range of masses of
O-helium with the minimal required central density of O-helium dark matter at $m_O = 1.25 \TeV$. For smaller or larger values of $m_o$ one needs larger central density to provide effective excitation of O-helium in collisions. Current analysis favors lowest values of central dark matter density, making possible O-helium explanation for this excess only for a narrow window around this minimal value.

In a two-component dark atom model, based on Walking Technicolor, a
sparse WIMP-like component of atom-like state, made of positive and negative doubly charged techniparticles, is present together with the dominant OHe dark atoms. Decays of doubly positive charged techniparticles to pairs of same-sign leptons can explain \cite{AHEP} the excess of high-energy cosmic-ray
positrons, found in PAMELA and AMS02 experiments\cite{PAMELA,AMS-2old,AMS2,AMS1}. Since even pure
lepton decay channels are inevitably accompanied by gamma radiation the
important constraint on this model follows from the measurement of cosmic
gamma ray background in FERMI/LAT experiment\cite{FERMI}. The multi-parameter
analysis of decaying dark atom constituent model determines the maximal model independent value of the mass of decaying
+2 charge particle, at which this explanation is possible $$m_O<1 TeV.$$

One should take into account that even in this range hypothesis on decaying composite dark matter, distributed in the galactic halo, can lead according to  \cite{kb} to gamma ray flux exceeding the measurement by FERMI/LAT. It can make more attractive interpretation of these data by an astrophysical pulsar local source\cite{semikoz} or by some local source of dark matter annihilation or decay. 

Experimental probes for OHe dark matter at the LHC strongly differ from the usual way of search for dark matter at acelerators, involving missed energy and momentum detection. Pending on the nature of the double charge constituents it may be search for new stable $U$-hadrons (heavy stable hadrons that appear in the result of production of $U \bar U$ pair) or search for stable double charged lepton-like particles. In the first case there are applicable constraints from the search for supersymmetric R-hadrons, having similar experimental signatures and giving the minimal mass for $UUU$ close to 3 TeV. It excludes OHe interpretation of the cosmic positron anomalies in terms of heavy quark cluster constituents of OHe. 

The possibility to interpret cosmic positron anomalies in terms of OHe costituents that appear in the experiments as stable lepton-like double charged particles is also close to complete test. 
The~ATLAS and CMS collaborations at the LHC are searching for the~double charged particles since 2011 \cite{Aad:2013pqd,Aad:2015oga,Chatrchyan:2013oca}. The~most stringent results achieved so far exclude the~existence of such particles up to their mass of $680$~\GeV{}. This value was obtained by both ATLAS and CMS collaborations independently. It is expected that if these two collaborations combine their independently gathered statistics of LHC Run 2 (2015--2018), the~lower mass limit of double charged particles could reach the~level of about $1.3$~\TeV{}. It will make search for exotic long-living double charged particles an \textit{experimentum crucis} for interpretation of low and high energy positron anomalies by composite dark matter \cite{Bled2017,probes}.

The successful and self-consistent OHe scenario implies the existence of dipole Coulomb barrier, arising in OHe-nuclear interaction and supporting dominance of elastic OHe-nuclear scattering. This problem of nuclear physics of OHe remains the main open question of composite dark matter, which implies correct quantum mechanical solution \cite{CKW}. The lack of such a barrier and essential contribution of inelastic OHe-nucleus processes seem to lead to inevitable overproduction of anomalous isotopes \cite{CKW2}. The advantages of the qualitative picture of OHe scenario appeal to increase  the efforts to solve this problem.
\section{Primordial massive black hole clusters}\label{pbhs}
The standard cosmological model considers homogeneous and isotropic Universe as the result of inflation. The observed celestial objects and strong inhomogeneities are evolved from small primordial density fluctuations that are also originated from small fluctuations of the inflaton field. It seems that there is no room for strong primordial inhomogeneities in this picture. Moreover, the existence of large scale inhomogeneities at the scales $\gg 100$Mpc is excluded by the measured isotropy of CMB. 

However, BSM physics, predicting new fields and mechanisms of symmetry breaking, adds new elements in this simple scenario that provide the existence of strong primordial inhomogeneities. Such predictions are compatible with the observed global homogeneity and isotropy of the Universe, if the strongly inhomogeneous component $i$ with $(\delta \rho/\rho)_i \sim 1$ contributes into the total density $\rho_{tot}$ whithin the observed level of the large scale density fluctuations  $(\delta \rho/\rho) = \delta_0 \ll 1$. It implies either large scale inhomogeneities, suppressed by the small contribution of the component $i$ into the total density $\rho_i/\rho_{tot} \le \delta_0$, or inhomogeneities at small scales.

A simple example of an axion-like model with U(1) symmetry broken spontaneously and then explicitly illustrates these two possible forms of strong primordial inhomogeneities. 

In this model spontaneous U(1)
symmetry breaking is induced by the vacuum
expectation value \beq \label{psif} \langle \psi \rangle = f\eeq of a complex scalar
field \beq \label{psi} \Psi = \psi \exp{(i \theta)},\eeq having also explicit
symmetry breaking term in its potential \beq \label{lambda }V_{eb} = \Lambda^{4} (1 -
\cos{\theta})\eeq. 

If the first phase transion takes place after inflation at $T = f$ and $f \gg \Lambda$, the potential Eq. (\ref{lambda }) doesn't influence continuous degeneracy of vacua on $\theta$ and string network is formed, which is converted in a walls-surrounded-by-strings network, separating regions with discrete vacuum degeneracy $\theta_{vac}+0, 2\pi, ...$ after the second phase transition at $T=\Lambda$. The vacuum structure network is unstable and decays, but the energy density distribution of $\theta$ field oscillations is strongly inhomogeneous and retains the large scale structure of this network, as it was shown in the example of axion models in \cite{Sakharov2,kss,kss2}. To fit the observational constraints on the inhomogeneity at large scales the contribution into the total density of such structure, called \textit{archioles}, should be suppressed. It causes serious problem for CDM models, in which the dominant form of dark matter is explained by axions \cite{Sakharov2,kss,kss2}.

If the first phase transition takes place at the inflational stage and $f \gg \Lambda$, as it was considered in \cite{RKS}, there appears a valley
relative to values of phase in the field potential in this period.
Fluctuations of the phase $\theta$ along this valley, being of the
order of $\Delta \theta \sim H/(2\pi f)$ (here $H$ is the Hubble
parameter at inflational stage) change in the course of inflation
its initial value within the regions of smaller size. Owing to
such fluctuations, for the fixed value of $\theta_{60}$ in the
period of inflation with {\it e-folding} $N=60$ corresponding to
the part of the Universe within the modern cosmological horizon,
strong deviations from this value appear at smaller scales,
corresponding to later periods of inflation with $N < 60$. If
$\theta_{60} < \pi$, the fluctuations can move the value of
$\theta_{N}$ to $\theta_{N} > \pi$ in some regions of the
Universe. 

After reheating, when the Universe cools down to
temperature $T = \Lambda$ the phase transition to the true vacuum
states, corresponding to the minima of $V_{eb}$ takes place. For
$\theta_{N} < \pi$ the minimum of $V_{eb}$ is reached at
$\theta_{vac} = 0$, whereas in the regions with $\theta_{N} > \pi$
the true vacuum state corresponds to $\theta_{vac} = 2\pi$. For
$\theta_{60} < \pi$ in the bulk of the volume within the modern
cosmological horizon $\theta_{vac} = 0$. However, within this
volume there appear regions with $\theta_{vac} = 2\pi$. These
regions are surrounded by massive domain walls, formed at the
border between the two vacua. Since regions with $\theta_{vac} =
2\pi$ are confined, the domain walls are closed. After their size
equals the horizon, closed walls can collapse into black holes.

The mass range of formed BHs is constrained by fundamental
parameters of the model $f$ and $\Lambda$. The maximal BH mass is
determined by the condition that the wall does not dominate locally
before it enters the cosmological horizon. Otherwise, local wall
dominance leads to a superluminal $a \propto t^2$ expansion for the
corresponding region, separating it from the other part of the
Universe. This condition corresponds to the mass \cite{book2}\beq
\label{Mmax} M_{max} =
\frac{m_{pl}}{f}m_{pl}(\frac{m_{pl}}{\Lambda})^2.\eeq The minimal
mass follows from the condition that the gravitational radius of BH
exceeds the width of wall and it is equal to\cite{book2,KRS}\beq
\label{Mmin} M_{min} = f(\frac{m_{pl}}{\Lambda})^2.\eeq

This mechanism can lead to formation
of primordial black holes of a whatever large mass (up to the mass
of active galactic nuclei (AGNs) \cite{AGN,DER1}, see \cite{PBHrev} for the latest review). Such black
holes appear in the form of primordial black hole clusters,
exhibiting fractal distribution in space
\cite{book2,KRS,Khlopov:2004sc,PBHrev}. It can shed new light on the
problem of galaxy formation \cite{book2,DER1,PBHrev}.

Closed wall collapse leads to primordial GW spectrum, peaked at \beq
\label{nupeak}\nu_0=3\cdot 10^{11}(\Lambda/f){\rm Hz} \eeq with
energy density up to \beq \label{OmGW}\Omega_{GW} \approx
10^{-4}(f/m_{pl}).\eeq At $f \sim 10^{14}$GeV this primordial
gravitational wave background can reach $\Omega_{GW}\approx
10^{-9}.$ For the physically reasonable values of \beq
1<\Lambda<10^8{\rm GeV}\eeq the maximum of spectrum corresponds to
\beq 3\cdot 10^{-3}<\nu_0<3\cdot 10^{5}{\rm Hz}.\eeq In the range from tens to 
thousand Hz such background may be a challenge for Laser Interferometer Gravitational-Wave Observatory (LIGO) experiment. 

Another
profound signature of the considered scenario are gravitational wave
signals from merging of BHs in PBH cluster. Being in cluster, PBHs with the masses of tens $M_{\odot}$ form binaries much easier, than in the case of their random distribution. In this aspect detection of signals from binary BH coalescence in the gravitational wave experiments \cite{gw1,gw2,gw3,gw4,gw5} may be considered as a positive evidence for this scenario. Repeatedly detected signals localized in the same place would provide successive support in its favor \cite{PBHrev,pbhClusters}.

\section{Antihelium from antimatter stars in our Galaxy}\label{antiMatter}
Primordial strong inhomogeneities can also appear in the baryon
charge distribution. The appearance of antibaryon domains in the
baryon asymmetrical Universe, reflecting the inhomogeneity of
baryosynthesis, is the profound signature of such strong
inhomogeneity \cite{CKSZ}. On the example of the model of
spontaneous baryosynthesis (see \cite{Dolgov} for review) the
possibility for existence of antimatter domains, surviving to the
present time in inflationary Universe with inhomogeneous
baryosynthesis was revealed in \cite{KRS2}.

The mechanism of
spontaneous baryogenesis \cite{Dolgov,Dolgov2,Dolgov3} implies the existence of a complex scalar field \beq \chi =(f/\sqrt{2})\exp{(i \theta )}\eeq carrying the baryonic charge. The $U(1)$ symmetry, which corresponds to the baryon charge, is broken spontaneously and explicitly, similar to the case, considered in the previous Section \ref{pbhs}. The explicit
breakdown of $U(1)$ symmetry is caused by the phase-dependent term, given by Eq. (\ref{lambda }).

Baryon and lepton number violating interaction
of the field $\chi$ with matter fields can have the following
structure \cite{Dolgov} \beq\label{leptnumb} {\cal
L}=g\chi\bar QL+{\rm h.c.}, \eeq where fields $Q$ and $L$ represent
a heavy quark and lepton, coupled to the ordinary matter fields.

In the early Universe, at a time when the friction term, induced by
the Hubble constant, becomes comparable with the angular mass
$m_{\theta}=\frac{\Lambda^2}{f}$, the phase $\theta$ starts to
oscillate around the minima of the PNG potential and decays into
matter fields according to (\ref{leptnumb}). The coupling
(\ref{leptnumb}) gives rise to the following \cite{Dolgov}: as
the phase starts to roll down in the clockwise direction,  it preferentially creates
excess of baryons over antibaryons, while the opposite is true as it
starts to roll down in the opposite direction. If fluctuations of $\theta$ on inflational stage move its value above $\pi$ in some region, it starts to roll down anticlockwise and, simulataneously, there should appear a closed wall, separting this region. As we discussed in the previous Section, collapse of such wall leads to formation of black hole with the mass in the range, determined by $f$ and $\Lambda$.

The fate of such antimatter regions depends on their size. If the
physical size of some of them is larger than  the critical surviving size
$L_c=8h^2$ kpc~\cite{KRS2}, they survive annihilation with surrounding matter.

The possibility of formation of dense antistars within an extension of the Affleck-Dine scenario of baryogenesis and the strategies for their search were considered in \cite{Blinnikov:2014nea}.

Evolution of
sufficiently dense antimatter domains can lead to formation of
antimatter globular clusters \cite{GC}. The existence of such
cluster in the halo of our Galaxy should lead to the pollution of
the galactic halo by antiprotons. Their annihilation can reproduce
\cite{Golubkov} the observed galactic gamma background in the
range tens-hundreds MeV. The gamma background data put upper limit on the total mass of antimatter stars.  The prediction of antihelium component of
cosmic rays \cite{ANTIHE}, as well as
of antimatter meteorites \cite{ANTIME} provides the direct
experimental test for this hypothesis. In the mechanism of spontaneous baryosynthesis there appears an interesting possibility of PBH, associated with dense antimatter domain, and the observational constraints on presence of massive black holes in globular clusters put constraints on the parameters $f$ and $\Lambda$.

Cosmic antihelium flux is a well motivated stable signature of antimatter stars in our Galaxy \cite{GC,ANTIHE}. Antimatter nucleosynthesis produces antihelium 4 as the most abudant (after antiprotons) primordial element. Antimatter stellar nucleosynthesis increases its primordial abundance. Heavy antinuclei, released in anti-Supernova explosions, annihilate with interstellar gas (dominantly hydrogen) and give rise to multiple antihelium fragments in the result of annihilation \cite{GC,ANTIHE}. Propagation of antihelium-4 in the matter gas is also accompanied by its annihilation, in which about 25\% of events give fragments of antihelium-3, either directly or after antitritium decay. 

The minimal mass of antimatter globular cluster is determined by the condition of the sufficient survival size of antimatter domain, corresponding to $10^3 M_{odot}$. It leads to a minimal cosmic antihelium flux accessible to searches for cosmic ray antihelium in AMS02 experiment.

Possible evidences for positive results of these searches continuously appear in the presentations by the AMS collaboration \cite{amsAHe3,amsAHe}. To the present time there are about ten clear candidates for antihelium-3 and two events that may be interpreted as antihelium-4. These results need further analysis and confirmation. It is expected that more statistics and the 5$\sigma$ result will be available to 2024. It would be interesting to check whether significant amount of matter in the aperture of AMS02 detector hinder antihelium-4, but increase antihelium-3 fraction in the result of antihelium-4 annihilation with matter. Confirmation of antihelium-3 events that cannot be explained as secondary from cosmic ray interactions \cite{poulin} would favor antimatter globular cluster hypothesis appealing to its detailed analysis.

\section{Conclisions}\label{conclusions}
 To conclude, the BSM physical basis leads to nontrivial features in cosmological scenario and the current exprimental progress probably gives evidences favoring their existence. However these evidences need further confirmation as well more theoretical work is needed to confront these predicted features with the experimental data. 
 
 Indeed, even in the simplest case of OHe dark atoms the open problem of OHe interaction with nuclei hinders a possible OHe solution for the puzzles of direct dark matter searches, On the other hand, indirect effects of OHe dark matter can explain anomalies of low and high energy cosmic positrons only for masses of hypothetical stable double charged within the reach of the search for such particles at the LHC. It opens new line of accelerator probes for dark matter.
 
Prediction of primordial strong inhomogeneities in the distribution of
total, dark and baryonic matter in the Universe is the new
important phenomenon of cosmological models, based on BSM physics with hierarchy of symmetry breaking.
The current progress in detection of gravitational waves and cosmic antimatter nuclei is probably approaching confirmation for the corresponding nonstandard cosmological scenarios.

Here we have given examples of nontrivial cosmological consequences coming from some minimal extensions of particle Standard model, involving prediction of extra stable double charged particles or additional global U(1) symmetry. One can expect much richer set of predictions in a more extensive theoretical framework of BSM physics and the platform of Bled Workshops will provide a proper place for extensive nonformal discussion of such a rich new physics and its cosmological impact.
\section*{Acknowledgements}
The work was supported by grant of Russian Science Foundation (project N-18-12-00213).


\end{document}